\documentclass[prd,onecolumn,nofootinbib]{revtex4} 
\usepackage{latexsym}
\usepackage{amsthm}
\usepackage{ulem}
\usepackage{amsmath}
\usepackage{enumitem}
\usepackage{graphicx}  
\usepackage{bm}        
\usepackage{amssymb}   
\usepackage{hyperref}
\usepackage{relsize}
\usepackage{color}
\newcommand{\identity}{{\rlap{1} \hskip 1.6pt \hbox{1}}}
\begin{document}
\title{Orientability of loop processes in relative locality}
\author{Lin-Qing Chen}
\affiliation{Perimeter Institute for Theoretical Physics, Waterloo, Ontario N2L 2Y5, Canada}
\begin{abstract} 
Inspired by recent results of unusual properties of loop processes in relative locality, we introduce a way to classify loops in the case of $\kappa$-Poincar\'e momentum space. We show that the notion of orientability is deeply connected to a few essential properties of loop processes. Nonorientable loops have “effective curvature”, which explicitly breaks translation symmetry, and can lead to a breaking of causality and  global momentum conservation.  Orientable loops, on the other hand, are ``flat."  Causality and global momentum conservation are all well preserved in this kind of loops. We comment that the nontrivial classical loops in relative locality might be understood as dual effects from general relativity, and some physical implications are discussed. 
\end{abstract}
\maketitle
\tableofcontents
\section{Introduction}
Relative locality  (RL) is a proposal for describing the Planck scale modification to relativistic dynamics in a particular limit in which $\hbar \rightarrow 0$, $G_N \rightarrow 0$, while $M_p=\sqrt{\hbar c/G_N}$ is fixed \cite{AmelinoCamelia:2011bm,  AmelinoCamelia:2011pe, Kowalski-Glikman:2013rxa}.  The insight behind relative locality is that we never directly measure the spacetime which we postulate. Instead, the fundamental measurements are quanta of the energy, momenta, and times of events.  Hence it is of interest to study the dynamics based on taking momentum space as primary, with spacetime being an emergent concept. The idea that momentum space should have a non-trivial geometry when quantum gravity effects are taken into account was originally proposed by Max Born \cite{Born}. One of the insights of quantum mechanics is an equivalence between space and momentum space, which is the so-called Born reciprocity.  Thus allowing the momentum space to have a nontrivial geometry is a natural way to reconcile gravity with quantum mechanics from this viewpoint \cite{AmelinoCamelia:2011pe},\cite{Born2}.  Relative locality could be treated as a dual gravity theory in the sense that in RL, momentum space is the basis manifold, while spacetime is the cotangent space (flat and linear) attached at each momentum point, which is the opposite of general relativity. In GR, the phase space of a point-like particle is the cotangent bundle of the spacetime manifold $\mathcal {M} : \Gamma^{GR}=T^*(\mathcal{M})$, while in RL, the phase space is the cotangent bundle of the momentum manifold $\mathcal {P} : \Gamma^{RL}=T^*(\mathcal{P})$  \cite{AmelinoCamelia:2011pe}.  

Loop processes in relative locality give new phenomenological predictions that can possibly be tested by experiments \cite{Freidel:2011mt, CamoesdeOliveira:2011gy, AmelinoCamelia:2011nt}, and also some unusual  phenomena at very high energies that have not been fully understood \cite{Chen:2012fu, Banburski:2013wxa}. Previous research has shown that when the geometry of momentum space is taken to be $\kappa$-Poincar\'e \cite{Majid:1994cy, Gubitosi:2011ej, AmelinoCamelia:2011nt},  a lot of loop processes would break causality and global momentum conservation, with the strange property of ``x dependence" , and these weird phenomena are only apparent when the energy scale is close to the Planck energy. \cite{Chen:2012fu, Banburski:2013wxa}. They are important properties that mark the departure of relative locality from usual special-relativistic dynamics.  It is then natural to ask what kinds of loops will have these unusual properties.  Is there any criterion that we can use to predict whether these strange phenomena will occur or not, without having to carry out a full calculation? Are there any deep connections between the violation of causality, x dependence, and global momenta nonconservation in relative locality, and what are the essential reasons for them to occur?

In this paper we are going to answer these questions and also give a simple criterion for predicting what kind of loops will exhibit these strange properties. It turns out that the orientability of loop processes is an essential concept. Nonorientable loops have ``effective curvature'' caused by combinations of nonlinear interactions. This effective curvature will make a loop configuration dependent. It can be understood as dual to the spacetime curvature in general relativity, and  also similarly a loop in curved spacetime is momentum dependent. In relative locality this kind of loops can possibly break causality and does not have global momentum conservation in general.  On the other hand, orientable loops have less novel effects and also make less trouble. We show that the orientable loops are flat and x independent.  Causality and global momenta conservation are very well preserved for this class of loops. In this way, connections between these properties are established and the essential reasons for unusual phenomena to occur becomes very apparent in a precise sense.  We then discuss the physical implications and prospects for novel quantum gravity phenomenology.

\section{Preparation}
\subsection{Dynamics of particles in relative locality}
In relative locality, we assume the momentum space to be a manifold $\mathcal{P}$ with a metric and a connection. The manifold $\mathcal{P}$ can possibly have torsion, curvature, or nonmetricity in general. The rest mass of a particle with momentum $p$ is given by the geodesic distance $\mathcal{D}(p)$ between $p$ and the origin of momentum space\cite{AmelinoCamelia:2011bm}. The addition rule of momenta between two particles is defined as a map, 
\begin{equation*}
\begin{split}
\oplus: \mathcal{P} \times \mathcal{P} &\rightarrow \mathcal{P}\\
(p,q) & \mapsto p\oplus q
\end{split}
\end{equation*}
which has an inverse $\ominus$, such that $(\ominus p) \oplus p = 0$. In an n-particle interaction, the momentum conservation equation is represented by, for example $\mathcal K (p^1, p^2...p^n)=(p^1 \ominus p^2) \ldots \oplus p^n  \equiv 0$, in which $\ominus p$ corresponds to outgoing particles' momenta. Since in relative locality, the nonlinear momenta addtion is not assumed to be commutative or associative, different orders of addition give different momenta conservation equations, which reflect the microscopic causal order of an interaction vertex \cite{AmelinoCamelia:2011bm}.

The dynamics of classical particles in relative locality is described by the action \cite{AmelinoCamelia:2011bm}:
\begin{equation}
S=\sum_J \! \int _{s_i} ^{s_j} {\!ds(x_J^a \dot{p}^J_a +\! \mathcal N_J \mathcal C^J (p^J))}  + \sum_i  \mathcal K^i_a(p^J(s_i)) z_i^a
\end{equation}
The index $J$ labels different particles, and the indices $i,j$ label different interaction vertices. In the free part of the action, $x_J$ are \textit{Hamiltonian spacetime coordinates} which are defined as being canonically conjugate to $p^J_a$:  $\{p^J_b, x_I^a\} = \delta^a_b \delta^J_I$ and $x_J^a \in T^*_{p^J}$, which means each particle lives in its own ``spacetime''; $s$ is an affine (time) parameter along the trajectory of the particle on $T^*_{p^J}$  and an interaction labeled by $i$ happens at $s_i$ for each particle. The mass-shell condition
$\mathcal C^J(p)\equiv \mathcal D^2(p^J)-m^2_J $ is imposed by the Lagrange multiplier $\mathcal N_J$. The interaction part of the action is a Lagrange multiplier times the conservation of momenta 
$\mathcal K_a (p^1, p^2...)\equiv 0$.  By varying the action we get the equations of motion, two of which are nontrivial:
\vspace{-1 mm}
\begin{eqnarray}
\dot{x}^a_J =\mathcal N_J \frac{\partial \mathcal C(k^J)}{\partial k_a^J}
\label{four velocity} \\
x^a_J(s_i)=\pm z^b_i \frac{\partial \mathcal K^i_b}{\partial k^J_a}
\label{end points}
\vspace{-1 mm}
\end{eqnarray}
$\pm$ indicates an incoming/outgoing particle respectively. Equation (\ref{four velocity}) tells us how free particles propagate on the cotangent space (``Hamiltonian spacetime" $T^*_p$) of their momentum space.  Equation (\ref{end points}) comes from the variation of the boundary terms of the action, and it describes how interaction events connect the starting or ending points of the worldlines on each particles' cotangent spaces $T^*_p$. In Eq.(\ref{end points})  $\partial \mathcal K^i/ \partial k^J$ is a transport operator $T^*_0 \rightarrow T^*_{k^J}$ transporting covectors between different cotangent spaces, and thus it connects different particles' own ``spacetimes" through an interaction vertex.  When the geometry of momentum space is a linear vector space (which is the assumption in current physics), the relationships given by Eq.(\ref{end points}) degenerate to $x_J^a(s_i)= z^a_i $, which results in the emergence of a Minkowski spacetime that all the particles live in. 

Note that the transport operator $\partial \mathcal K^i/ \partial k^J$ can be quite complicated when there are many particles involved in the interaction. For notation convenience, we define the left and right transport operators\cite{Freidel:2011mt}, 
\begin{equation}
\frac{\partial (p\oplus q)}{\partial q}\Big|_{p\oplus q=k} :=U^q_{k}, \ \ \ \frac{\partial (p\oplus q)}{\partial p} \Big |_{p \oplus q =k}:=V^p_{k}
\label{UV}
\end{equation}
They transport covectors from $T^*_k \rightarrow T^*_{q}$ and $T^*_k \rightarrow T^*_{p}$ respectively. We also define the inverse operator $I_p$ which sends covectors from $T^*_{\ominus p} \rightarrow T^*_{p}$ \cite{Freidel:2011mt},
\begin{equation}
\frac{\partial{\ominus p}}{\partial {p}} :=I_p
\end{equation}
We introduce them here because later we will show that the transport operator $\partial \mathcal K^i/ \partial k^J$ can always be written as a product of a few $U$'s, $V$'s, and $I$'s \cite{Freidel:2011mt}. Thus they are building blocks of the transport operator. In the next subsection, we will study the properties of these transport operators in a specific geometry of momentum space.

\subsection{Properties of transport operators in $\kappa$-Poincar\'e momentum space}
In this section we set up a few properties of addition-induced transport operators in $\kappa$-Poincar\'e momentum space, which  is a nonlinear momentum space with a de Sitter metric, and a torsionful connection. It has vanishing curvature but nonzero torsion and nonmetricity \cite{Gubitosi:2011ej, AmelinoCamelia:2011nt}.  The symmetry of $\kappa$-Poincar\'e momentum space originally comes from a dimensionful deformation of the Poincar\'e group into a Hopf algebra\cite{Majid:1994cy}.  The deformation parameter $\kappa$ is assumed to be the Planck scale. The addition rule of momenta is noncommutative,
\begin{equation}
\begin{split}
(p\oplus q)_0 &=p_0+q_0\\
(p\oplus q)_i &=p_i+ e^{-p_0/\kappa} q_i 
\label{addtionrule}
\end{split}
\end{equation}
One of the key properties of momenta addition is associativity, which comes from the coassociativity axiom of Hopf algebras, and can be checked by straightforward calculation.
The associativity immediately implies the left inverse property and the right inverse property,
\begin{equation}
 (\ominus p)\oplus(p\oplus q)=q, \ \ \ \ \ \ \ \ (q\oplus p)\ominus p=q
\label{ip}
\end{equation}
Differentiating the above equations gives us the following very useful relations between left and right transport operators\cite{Freidel:2013xx}:
\begin{equation}
\textnormal{Left}:\ U^{p\oplus q}_{q} \cdot V^{p}_{p\oplus q} + V^{\ominus p}_q \cdot I_p=0,\ \ \ \ \ \textnormal{Right}:\ V^{q\oplus p} _q \cdot U^{p}_{q\oplus p} + U^{\ominus p}_q\cdot I_p=0
\label{lr}
\end{equation}
By setting the momenta $q=0$ in the above Eq. (\ref{lr}), we can get the following relationships of the transport operators between the cotangent space of some momentum $p$ and the origin:
\begin{equation}
 V^{\ominus p}_0 \cdot I_p=-U^p_0,\ \ \ \ \ \ U^{\ominus p}_0 \cdot I_p=-V^p_0
\label{sip}
\end{equation}
The left  and right inverse properties (\ref{ip})  imply that  the inverses of the left and right transport operators are simply
$[U^p_q]^{-1} = U^q_p$ and $[V^p_q]^{-1}= V^q_p $. Note that the operators $U^p_q, V^p_q$ are always invertible when p is in the composition of momenta q.\\

Using the associativity, we can also obtain the chain rule for multiplication of transport operators:
\begin{equation}
U^p_k \cdot U^k_q = U^k_q \cdot U^p_k=U^p_q, \ \ \ \ \ V^p_k \cdot V^k_q = V^k_q \cdot V^p_k=V^p_q
\label{chain rule}
\end{equation}

Note importantly that when the momentum space has torsion as in the $\kappa$-Poincar\'e momentum space, there is no chain rule for the product of a mixture of $U$ and $V$, for example $U^p_q \neq (U^p_k \cdot V^k_q ) \neq V^p_q$. This will have significant consequences for the results in later sections. 

Applying the left and right inverse properties together with the chain rule (\ref{chain rule}), we have equations which turn out to be useful for simplifying expressions,
\begin{equation}
 U^{p\oplus q}_{q} \cdot V^{p}_{p\oplus q}=V^0_q \cdot U^p_0, \ \ \ \ \  V^{q\oplus p} _q \cdot U^{p}_{q\oplus p}=U^0_q \cdot V^p_0,
\end{equation}

We need to emphasize here that the above properties are not only valid for $\kappa$-Poincar\'e momentum space, but also for all kinds of momentum space geometries in which the addition rule of momenta is associative.

Specifically for $\kappa$-Poincar\'e momentum space,  a nice property of the left transport operator is:
\begin{equation}
U^{p\oplus q}_q=U^p_0 ,\ \ \textnormal{ or equivalently}\ \ U^p_q=U^{p\ominus q}_0
\label{left}
\end{equation}
which is not true for right transport operators. Taking the above relation (\ref{left}) together with the chain rule (\ref{chain rule}), we have another property:
\begin{equation}
U^p_0\cdot U^q_0=U^{p\oplus q}_0=U^{q\oplus p}_0
\end{equation}
We will need to use the above mathematical tools in the later sections.

\subsection{Causal relationships and Event-nets}
In general relativity, the causal relations between events are expressed as geometrical relations between points on the spacetime manifold. In the theory of relative locality,  the structure of spacetime has already radically changed. Particles do not live in a universal spacetime unless the momentum space is taken to be linear.  However, the notions of events and the existence of null or timelike propagating particles connecting events are still not changed. We define event i to be in the {\textit{causal past}} of event j:  $i\!\prec j$ if  in the process that is being considered, there exists a sequence of events $i, i+1, ... i+n, j, n\geq 0$ s.t. from each event there exists an outgoing free-propagating particle coming to the next event. Similarly we can define the {\textit{causal future}} $j\!\succ \!i$. \cite{Chen:2012fu}

An event-net is a directed graph \cite{Freidel:2013xx}, denoted as $\Gamma=(\{\mathcal{K}_i\},\{p^I_{i,j}, \tau^I_{i,j}\})$, representing the causal relationship of a piece of history of physical processes in a classical regime. An event (labeled by $i$) is represented by a node, which is associated with a conservation of momenta  $\mathcal{K}_i \equiv 0$. Particles freely propagate between events. To each arc, we associate a momentum $p^I_{i,j}$ and a proper time $\tau^I_{i,j}$. Index $I$ is used to distinguish different particles that connect two events $i$ and $j$. We can omit this index when it is obvious. Note that $p^I_{i,j}=\ominus p^I_{j,i}$. The whole event-net is not a graph that can necessarily be embedded in a universal spacetime, rather it is a representation of causal relationships. A novel physical point of view is that observers have to be included inside the event-net. An observer can just measure the part of information of the events that connect with him on the graph. Just as in the real world, observers have very limited information accessible to them.

\begin{figure}[h]
  \centering
    \includegraphics[width=0.8\textwidth]{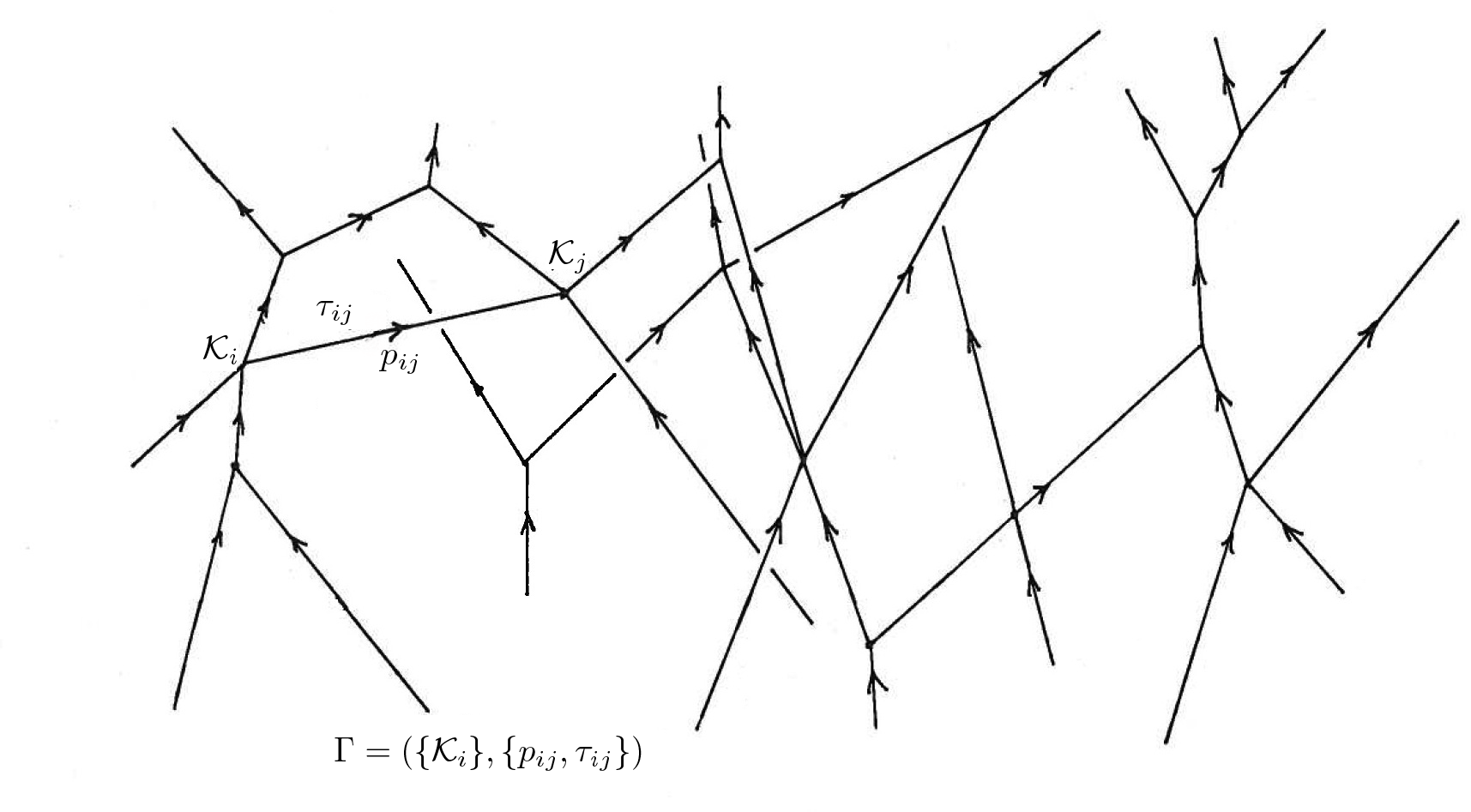}
\vspace{-5 mm}
	\caption{An example of a piece of event-net.}
\label{event-net}
\vspace{-3 mm}
\end{figure}

\subsection{Loop-closure condition in relative locality}
\begin{figure}[h]
  \centering
    \includegraphics[width=0.3\textwidth]{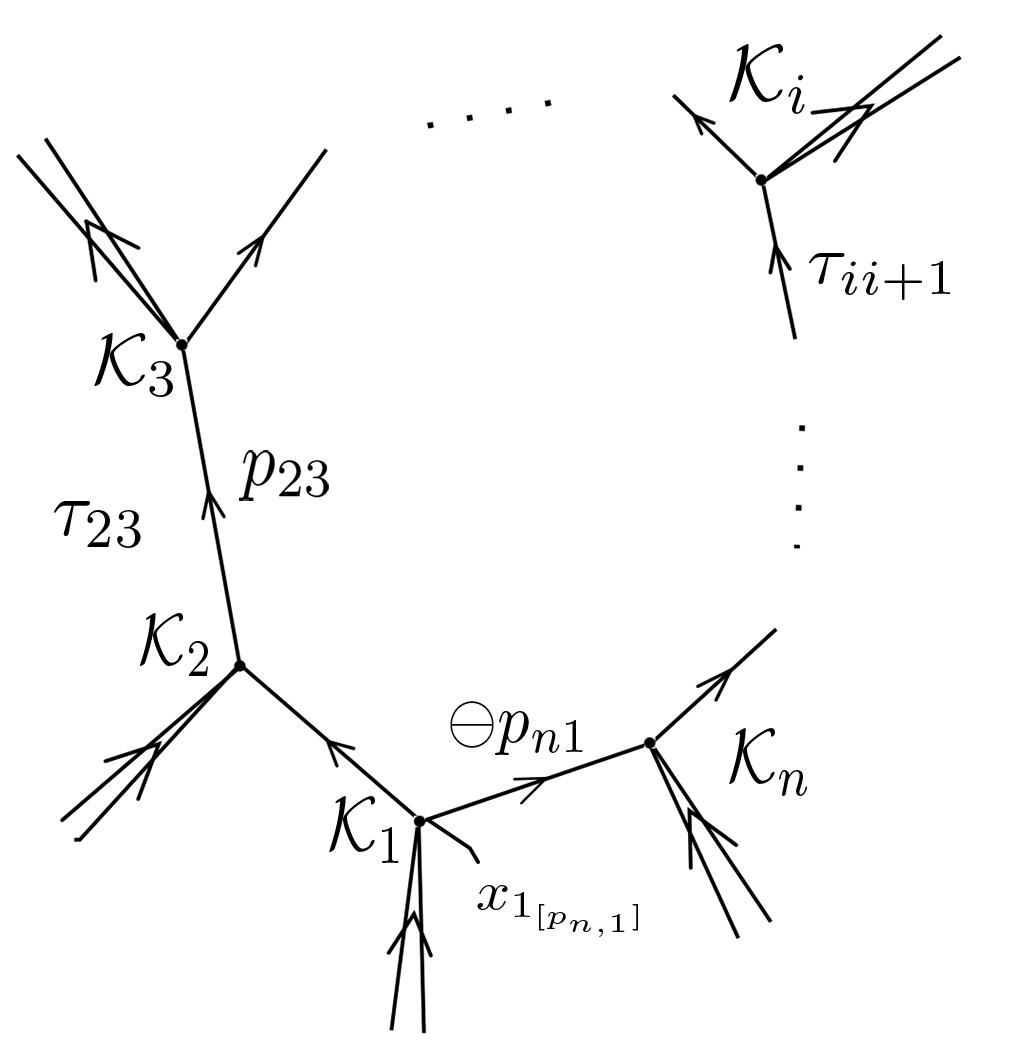}
\vspace{-5 mm}
	\caption{A loop with n vertices}
\label{a n vertices loop}
\vspace{-3 mm}
\end{figure}

Consider a loop with n vertices, in which each node is associated with an equation of momenta conservation $\mathcal{K}_1, \mathcal{K}_2 ... \mathcal{K}_n \equiv 0$.  Fig.[\ref{a n vertices loop}] shows the event-net of this loop.

\begin{figure}[h]
  \centering
    \includegraphics[width=0.7 \textwidth]{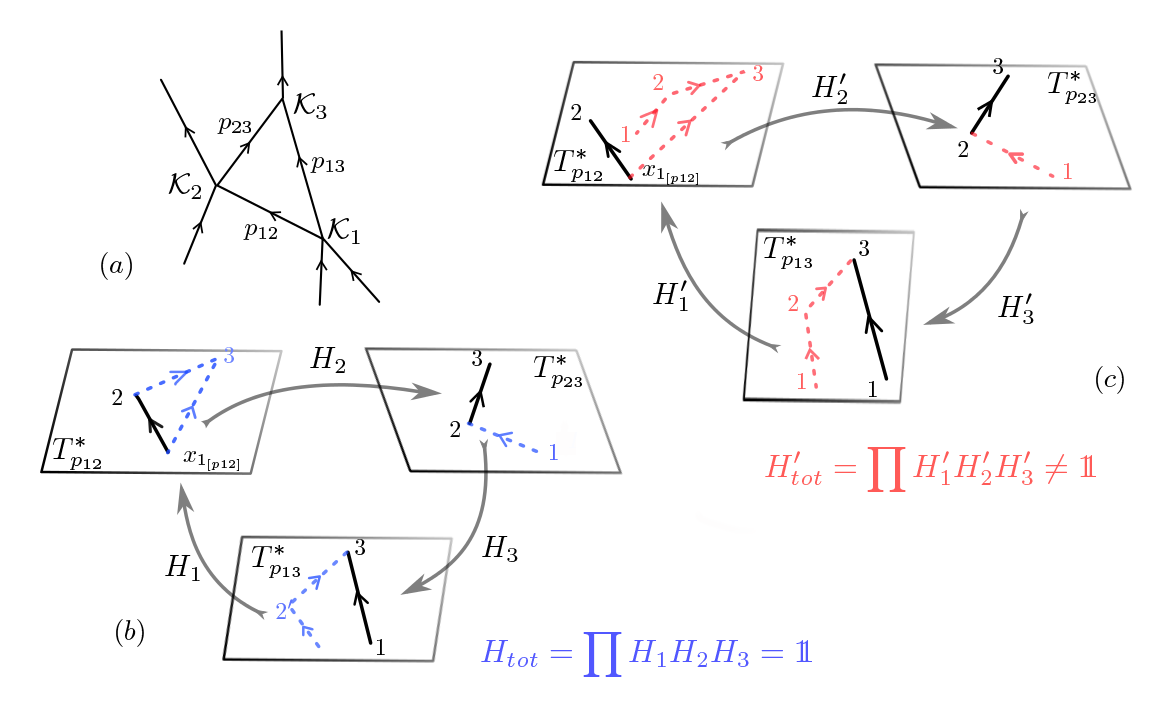}
	\caption{Comparison between a flat loop (b) and a loop with effective curvature (c). Both of them have the same event-net (a).}
\label{big}
\vspace{0 mm}
\end{figure}

First, we define a {\textit {vertex transport operator}} on each vertex $\mathcal{K}_i$ as 
\begin{equation}
H_{i} :=\left(\frac{\partial {K}_i}{\partial {p_{i-1,i}}} \right)^{-1}  \left(-\frac{\partial {K}_{i}}{\partial {p_{i,i+1}}} \right) 
\label{vto}
\end{equation}
It maps covectors from $T^*_{p_{i-1,i}} $to $T^*_{0} $ and then to $T^*_{p_{i,i+1}}$.  We can understand this vertex transport operator $H_{i}$ as sending the end point of the worldline of particle $p_{i-1,i}$ to the starting point of the worldline of particle $p_{i,i+1}$  in its ``Hamiltonian spacetime" $T^*_{p_{i,i+1}}$, see Fig [\ref{a n vertices loop},\ref{big}]. We use $x_{1_{[ p_{n,1} ] }}\in T^*_{p_{n,1}}$to label the end point of a particle $p_{n,1}$'s worldline which corresponds to the interaction vertex $\mathcal{K}_1$.

By imposing the equations of motion (\ref{four velocity}, \ref{end points}) around the loop, we get
\begin{equation}
x_{1_{[ p_{n,1} ] }}\cdot \left[ \identity-\prod_{i=1}^{n} H_{i}\right]= \sum_{i=1}^n \tau_{i,i+1} v_{i,i+1}   \prod_{i<j \leq n} H_{j} 
\label{closure}
\end{equation}
Note that we use the boundary condition in which $i=n+1 :=1$. Eq. \ref{closure} is the condition which has to be satisfied for the loop to be a solution of the theory. Equivalent conditions can be constructed in terms of the ending/starting point of any particles' worldline, and the equation will be the same up to  a cyclic permutation of a product of $H_i$s.

We define $H_{tot} := \prod_{i=1}^{n} H_{i}$ for shorthand notation. Notice that on the left-hand side of the equation, $[H_{tot} -\identity]$ measures the difference of a covector after transporting around the whole loop and coming back to the same cotangent space. It describes an {\textit{effective curvature}} in the loop, which is caused by a combination of nonlinear interactions. Note that this is not a curvature in the sense of Riemannian geometry, as $H_{tot}$ is not usually a holonomy for parallel transport.  If the effective curvature is nonzero, it imposes a constraint on the cotangent space $T^*_{p}$ such that a loop process can only happen at specific regions on each particles' ``Hamiltonian spacetime". In \cite{Chen:2012fu} this phenomena was termed x-dependence, which is an explicit breaking of translation symmetry on the cotangent spaces. Let us compare this to the more familiar case of loops in general relativity. There we can form closed loops by considering different geodesics, which are generated by four-velocities (and hence momenta). Hence the holonomy around a loop in curved spacetime is momentum dependent. Since relative locality is considered a dual picture of general relativity \cite{AmelinoCamelia:2011pe}, we can expect some loops to be configuration dependent.  It is interesting that  in relative locality spacetimes are flat, but since momentum space is nonlinear and there is no universal spacetime, we have an effective curvature in the loop processes. Now from (\ref{closure}) we know that 
\begin{equation}
x-independent  \Longleftrightarrow  H_{tot} := \prod_{i=1}^{n} H_{i}=\identity 
\label{condition}
\end{equation} 
The loop is x independent only when the chain product of transport operators around the whole loop is identity, i.e. the loop is flat. This is an important property that marks the departure of relative locality from usual special-relativistic dynamics, noting that the Newton constant $G_N \rightarrow 0$ in RL.

\vspace{3 mm}
The following figures show the difference between x-dependent loops and x-independent loops when we consider the transport of edges between particles' different ``Hamiltonian spacetimes". Consider a loop with three vertices, as in Fig.[\ref{big}](a). If the loop is flat, i.e. the transport operator along the whole loop is identity, every edge comes back to itself after transporting through other particles' cotangent spaces along the loop; see Fig.[\ref{big}](b).  If the loop has effective curvature, i.e. the transport operator along the whole loop is not identity, edges do not come back to themselves after transporting through other cotangent spaces along the whole loop; see Fig.[\ref{big}](c).

Note here that we did not make an equivalence between  x independence and cotangent space translation symmetry, a restricted definition of which was studied in \cite{AmelinoCamelia:2011nt}. Translation symmetry is very non-trivial for loop processes in which each particle lives in different cotangent spaces. For example consider a flat (x-independent) loop in the above fig.[\ref{big}] (a,b),  the whole process is invariant under the translation $x_* \rightarrow x_*+\Delta x_*$, where $\Delta x_*$ satisfy
\begin{equation}
\Delta x_{1_{[p_{12}]}}\cdot H_2 =\Delta x_{2_{[p_{23}]}} =\Delta x_{3_{[p_{13}]}}\cdot H_3^{-1}
\end{equation}

\subsection{Orientability of loop processes}
 As we have already discussed, the addition of momenta is not commutative in $\kappa$-Poincar\'e momentum space, and hence different orders of composition in the momenta conservation equations lead to different physical effects. Some orders in the vertices lead to strange and novel phenomenology in loop processes, such as effective curvature and x dependence, violation of causality and global momenta conservation, and the effect are obvious when the energy scale is close to Planck energy\cite{Chen:2012fu, Banburski:2013wxa}. It would be very useful to have a criterion for classification of loop processes, such that we can predict whether the strange effects will occur or not in a loop process just by simply looking at the form of interaction vertices, without having to carry out a full calculation. In the rest of the paper we will show that indeed such a classification exists and it is surprisingly simple. The criterion turns out to be orientability.

Consider a single loop -- at each vertex there are two internal momenta that form the boundary of the loop, and a few external momenta. If in the momenta conservation equation the external momenta are neighbours with each other without being separated by internal momenta, we just group the external ones to be a whole, for the convenience of discussion. We can have a notion of orientation by embedding the event-net of the loop in an oriented two-dimensional surface\footnote{We are considering single loops -- they can always be embeded in a two-dimensional plane.  We will discuss non-planar processes in the  Sec. III D.}.  Note that this two-dimensional surface is not anything physical like spacetime, it is just a mathematical abstraction.  For each vertex of the loop, the orientation of the vertex can be defined by the order of momenta addition.  We say it is {\textit{left-oriented vertex}} if the addition order of the three momenta -- two internal ones and the grouped external ones -- forms left-hand order on the oriented two-dimensional surface.  Similarly we can define the {\textit{right-oriented vertex}}.  If the external momenta cannot be grouped, we say that the vertex is {\textit{nonorientable}}.   We will justify the above naming in the next section by showing that for a left-oriented vertex, the vertex transport operator $H$ (\ref{vto}) is actually a left transport operator $U$. Similarly for the right-oriented vertex.  Now consider the whole loop. Only if all the vertices in the loop have the same orientation, do we say that  {\textit{the loop is orientable}}.  If in a loop it happens that some vertices are nonorientable, or have opposite orientations, then it is  {\textit{a nonorientable loop}}.

\begin{figure}[h]
  \centering
    \includegraphics[width=0.85 \textwidth]{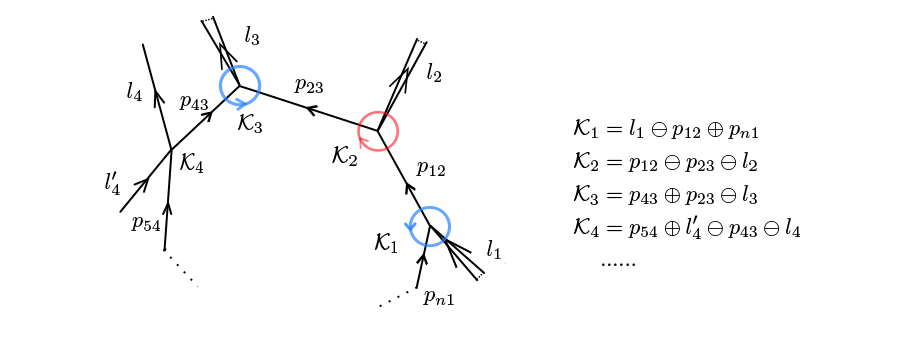}
\vspace{-5 mm}
	\caption{An example of a non-orientable loop}
\label{orientationex}
\vspace{-3 mm}
\end{figure}

Let us look at an example. In Fig.[\ref{orientationex}] below, $\mathcal {K}_1$, and $\mathcal {K}_3$ are right-oriented vertices, and $\mathcal {K}_2$ is left-oriented. $\mathcal{K}_4$  is a nonorientable vertex.  Hence the whole loop is not orientable.

\section{Implications of orientability}

\subsection{Orientability $\Leftrightarrow$ x independence/Flatness}
From the closure conditon (\ref{closure}) in the last section, we have shown that some loops may have effective curvature while some loops are flat.  It is a novel property that differentiates relative locality from special relativity. In this section, we will show that we can predict whether a loop is flat and x independent simply by looking at what kind of vertices the loop has.  The following is a proof that for single loop processes, a loop is flat and x independent if and only if it is orientable. 

Just as for a 3-vertex, there are 6 ways of writing momentum conservation equation in an associative momentum space. Thus for a whole loop with $n$ vertices, there are $6^n$ ways of writing the vertex equations in total. We will show that it is easy to dramatically simplify the analysis if we just want to look at the transport operator $H$. First we consider a vertex $\mathcal {K}$ with  $l_1$ and $l_2$ representing external momenta, and $p_1$ and $p_2$ representing two of the momenta which form the boundary of a loop.  If we permute the momenta in the vertex, in general, the different conservation laws can be written as:
\begin{equation}
\mathcal {K}_A \equiv p_1 \oplus l_1 \oplus p_2 \oplus l_2=0 \Rightarrow \mathcal {K}_{A}' \equiv l_1 \oplus p_2 \oplus l_2 \oplus p_1=0 \Rightarrow \mathcal {K}_{A}'' \equiv  p_2 \oplus l_2 \oplus p_1 \oplus l_1=0
\label{permutation}
\end{equation}
Though the above equations have the same set of solutions in terms of momenta, the transport operators are different:
\begin{equation*}
d_{p_1} \mathcal {K}_A \neq d_{p_1} \mathcal {K}_{A}' \neq d_{p_1} \mathcal {K}_{A}''
\end{equation*}
However, noticing that \cite{Freidel:2013xx},
\begin{equation*}
d_{p_1} \mathcal {K}_{A}' =U^{p_1}_0 V^0_{p_1} \cdot d_{p_1}\mathcal{K}_{A}, \ \ \ 
d_{p_2}\mathcal {K}_{A}' =U^{\ominus l_1}_0 V^{p_2}_{\ominus l_1} (U^{p_2\oplus l_2}_0 V^{p_2}_{p_2\oplus l_2})^{-1} \cdot d_{p_2}\mathcal {K}_{A}=U^{p_1}_0 V^0_{p_1} \cdot d_{p_2}\mathcal{K}_{A}
\end{equation*}
\begin{equation*}
d_{p_1} \mathcal {K}_{A}'' =U^{l_1}_0 V^0_{l_1} \cdot d_{p_1}\mathcal{K}'_{A}, \ \ \ 
d_{p_2}\mathcal {K}_{A}'' =U^{l_1}_0 V^0_{l_1} \cdot d_{p_2}\mathcal{K}'_{A}
\end{equation*}
From the above pattern, we see that the vertex transport operator (\ref{vto}) stays invariant under the permutation of the vertex:
\begin{equation}
H_A'=(d_{p_1} \mathcal{K}_{A}')^{-1} (-d_{p_2} \mathcal{K}_{A}')  =H_A=H_A''
\end{equation}
Thus for single loops, we actually just have to consider two forms of vertices: the left-oriented or the right-oriented, for the analysis of the vertex transport operator $H$. This is one of the advantages of introducing those concepts.  All the other forms of vertices can be obtained through permutations, which will give the same vertex transport operator.  We choose the two vertices we will study to be,
\begin{eqnarray}
\mathcal{K}_i=l_i\oplus p_{i-1,i}\ominus p_{i,i+1}
\label{conservationlaw1}\\
\mathcal{\tilde{K}}_i=l_i \ominus p_{i,i+1} \oplus p_{i-1,i}
\label{conservationlaw2}
\end{eqnarray}
where $p_{**}$  are the momenta  forming the boundary of the loop, and the indices $i,i+1$ indicate the assumed direction of momenta flow: from vertex $i$ to vertex $i+1$.  $l_i$ represents the total external momenta of the vertex in the loop. 
\begin{figure}[h]
  \centering
    \includegraphics[width=0.5\textwidth]{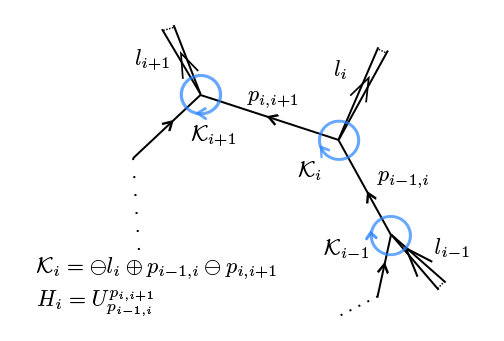}
\vspace{-5 mm}
	\caption{A left-oriented loop}
\label{An orientable loop}
\vspace{-3 mm}
\end{figure}

Now we consider an orientable loop with n vertices with the conservation laws on each vertex  given by (\ref{conservationlaw1}). The vertex transport operators are then
\begin{equation}
\begin{split}
H_i&=(d_{p_{i-1,i}} \mathcal{K}_{i})^{-1} (-d_{p_{i,i+1}} \mathcal{K}_{i}) \\
&=-(U^{\ominus l_i}_0 V^{p_{i-1,i}}_{\ominus l_i})^{-1} (U^{\ominus p_{i,i+1}}_0 I_{p_{i,i+1}})\\
&=U^{p_{i,i+1}}_{p_{i-1,i}}
\end{split}
\end{equation}
where we have used the chain rule (\ref{chain rule})  and the right inverse property (\ref{ip}) to get the final result.  For convenience we say that the loop is left-orientable because the vertex transport operator is the left one $U$ (\ref{UV}). By the chain rule, the transport operator of the whole loop is
\begin{equation}
H_{tot}=\prod_{i=1}^{n} U^{p_{i,i+1}}_{p_{i-1,i}} =\identity
\end{equation}
Thus the left-oriented loop is flat and x independent.  Fig.[\ref{An orientable loop}] is an example.

\vspace{3mm}
Similarly, for the other orientation (\ref{conservationlaw2}) we have
\begin{equation}
\tilde{H}_i = -( U^{p_{i-1,i}}_0)^{-1} U^{\ominus l_i}_0 V^{\ominus p_{i,i+1}}_{\ominus l_i} I_{p_{i,i+1}}=V^{ p_{i,i+1}}_{p_{i-1,i}}
\end{equation}
and the transport operator of the whole loop is $\tilde{H}_{tot}=\prod_{i=1}^{n} V^{p_{i,i+1}}_{p_{i-1,i}} =\identity$. This kind of loops is right-orientable. Thus we have just proved that an orientable loop has zero effective curvature and is x independent.
\\ \\
Now let us consider nonorientable single loops. There are actually just two kinds of them:
\begin{itemize}[label={$\cdot$}]
\item All the vertices in the loop are orientable, with different orientations. 
\item Some vertices in the loop are not orientable.
\end{itemize}
Let us start with the first case. The simplest example would be constructed by taking the left-oriented loop we were just considering in Fig.[\ref{An orientable loop}], and switching the orientation of one vertex in it. Let us say the left-oriented vertex $m$ in this loop becomes a right-oriented vertex and its momenta conservation equation is given by (\ref{conservationlaw2}). This leads to the following  transport operator $H'_{tot}$ around the loop
\begin{equation}
\begin{split}
H'_{tot}&=\prod_{i=1}^{m-1} U^{p_{i,i+1}}_{p_{i-1,i} } \cdot V^{ p_{m,m+1}}_{p_{m-1,m}} \cdot \prod_{i=m+1}^{n} U^{p_{i,i+1}}_{p_{i-1,i} }\\
&=U_{p_{n,1}}^{p_{m-1,m}}\cdot V^{ p_{m,m+1}}_{p_{m-1,m}} \cdot U_{ p_{m,m+1}}^{p_{n,1}}  
\end{split}
\end{equation}
If the momentum space has torsion, as is the case in $\kappa$-Poincar\'e  \cite{Gubitosi:2011ej}, we have
\begin{equation}
d_{p_{\mu}} \left[V^{q_\nu} _{(q\oplus p)_\rho}-U^{q_\nu} _{(p\oplus q)_\rho}\right] \Big|_{p,q=0} =T^{\mu\nu}_\rho =\frac{1}{\kappa} \delta^{[\mu}_0 \delta^{\nu]}_i \delta^i_\rho, \ \ \ \ i=1,2,3
\end{equation}
Hence $H'_{tot}$ cannot be identity:
\begin{equation}
H'_{tot}-\identity =U_{p_{n,1}}^{p_{m-1,m}}\cdot(V^{ p_{m,m+1}}_{p_{m-1,m}} -U^{ p_{m,m+1}}_{p_{m-1,m}})\cdot U_{ p_{m,m+1}}^{p_{n,1}} \neq 0
\end{equation}  
By switching more vertices to be the opposite orientation in the above loop, it is straightforward to see that the $H'_{tot}$ will never be identity until all the vertices turn out to have the same orientation again.
\\ \\
Let us now study the second case. In general, a nonorientable vertex is of the form:
\begin{equation}
\mathcal{K}_i=l_i\oplus p_{i-1,i}\oplus k_i\oplus(\ominus p_{i,i+1})
\label{nonorientable}
\end{equation}
where $l_i, k_i$ are two external momenta  at vertex $i$. The vertex transport operator of $\mathcal{K}_i$ for this case is,
\begin{equation}
H_i =-(U^{\ominus l_i}_0 V^{p_{i-1,i}}_{\ominus l_i})^{-1} (U^{\ominus p_{i,i+1}}_0 I_{p_{i,i+1}})=V^{p_{i-1,i}\oplus k_i}_{p_{i-1,i}} U^{p_{i,i+1}}_{p_{i-1,i}\oplus k_i}
\end{equation}
We can see that $H_i$ depends on $k_i$, i.e. one of the external momenta at vertex $i$ of the loop. Since  $k_i$ is independent from momenta at other vertices in the loop, a product chain of $H_i$ can not cancel away the dependence on $k_i$, thus $H_{tot}\neq \identity$.

Hence, nonorientable loops have effective curvature and are x dependent. In this section we have finally proved that orientability is the sufficient and necessary condition for x independence (flatness).

\subsection{Orientable loops preserve causality}
Previous work has shown that there are causal loops in relative locality \cite{Chen:2012fu}. Although it has not yet led to any known paradoxes,  a consistent unitary quantum theory for Relative Locality would be very nontrivial to construct due to the existence of those loops.  In this section we will show that by imposing orientability, causal loops will not form when momentum space is considered to be $\kappa$-Poincar\'e. We will also give an analysis of the reason why non-orientability can lead to causality violation.

We proved in the previous sections that for an orientable loop, the loop closure condition eq.(\ref{closure}) becomes
\begin{equation}
\sum_{i=1}^n \tau_{i,i+1} v_{i,i+1}^\mu  ( \prod_{i<j \leq n} H_{j} )_\mu^\nu =0
\label{causal}
\end{equation}
where the chain products of transport operators are
\begin{equation}
\prod_{i<j \leq n} H_{j}  =U^{p_{n,1}}_{p_{i,i+1}},\ \ \ \ or \ \ \ \ \prod_{i<j \leq n} H_{j}  =V^{p_{n,1}}_{p_{i,i+1}},
\end{equation}
which correspond to left and right orientation respectively.  To prove that orientable loops will not be causal loops, it suffices to show that the chain product of transport operators $ ( \prod_{i<j \leq n} H_{j} )_\mu^\nu$ can not reverse the direction of the future-pointing four velocity to past-pointing. More precisely, it should not change the sign of the zeroth component of $v^\mu$ after contraction if causality were to be preserved. 

As the left transport operator $U$ just has non-trivial elements on the diagonal  and
\begin{equation}
\left(U^{q}_{p}\right)^0_\mu= \delta^0_\mu,
\end{equation}
thus, for the loops which have left orientation, the zeroth component of eq. (\ref{causal}) is
\begin{equation}
\sum_{i=1}^n \tau_{i,i+1} v_{i,i+1}^\mu ( U^{p_{n,1}}_{p_{i,i+1}})_\mu^0 =\sum_{i=1}^n \tau_{i,i+1} v_{i,i+1}^0 =0
\end{equation}
It is immediate to see that the above equation does not have solutions when all the particles' four velocities are future-pointing, i.e. no causal loops can be formed. Thus left-oriented loops preserve causality.

For right-oriented loops, the causality preservation is less obvious. Unlike the form of $U$ above, the vertex transport operators of the right-oriented loops have non-trivial component in the first column, which in general have the form,
\begin{equation}
\left(V^q_p \right)^0_i=\frac{1}{\kappa} (q_i-p_i)
\end{equation}
Thus in this case, the zeroth component of each term in equation (\ref{causal}) is:
\begin{equation}
v_{i,i+1}^\mu \left( V^{p_{n,1}}_{p_{i,i+1}} \right)_\mu^0 = v_{i,i+1}^0+\frac{1}{\kappa}v_{i,i+1}^j \left[(p_{n,1})_j- (p_{i,i+1})_j\right]
\label{zero}
\end{equation}
For convenience, we use $p$ to represent $p_{i,i+1}$ and $q$ to represent $p_{n,1}$ in the above equation eq.(\ref{zero}). 

Now we need the explicit expression for the four velocity. In $\kappa$-Poincar\'e momentum space, the mass of a particle is given by the geodesic distance from the momentum point $p$ to the origin of momentum space\cite{Gubitosi:2011ej}:
\begin{equation}
m(p)= \kappa Arccosh(\cosh (p_0/\kappa)-e^{p_0/\kappa}  \mathsmaller{\mathsmaller{\sum}} p_i^2 /2\kappa^2)
\label{mass}
\end{equation}
Based on equation of motion (\ref{four velocity}),  we can get the four-velocity of a particle (with momentum $p$ and rest mass $m$) through taking derivative of the expression for mass(\ref{mass}):
\begin{equation}
v^0 = - \frac{e^{p_0/\kappa} \mathsmaller{\sum_i} p_i^2 -2\kappa^2 \sinh(p_0/\kappa)}{2 \kappa^2 \sinh(m/\kappa)}\ ,\ \ \ \ 
v^i =  \frac{ p^{i} e^{-p_0/\kappa} }{\kappa \sinh(m/\kappa)}
\label{4v}
\end{equation}
Plugging in (\ref{4v}) to (\ref{zero}), we get
\begin{equation}
v^\mu ( V^q_p )_\mu^0 = \frac{e^{p_0/\kappa}}{2 \kappa^2 Sinh (m/\kappa)} \underbrace{ \left[\kappa^2 -\kappa^2 e^{-2p_0/\kappa} +  \mathsmaller{\sum_i} p_i^2 -2  \mathsmaller{\sum_i} p_i q_i \right]}_\text{ $:= f(p_0, p_i, q_i)$}
\label{zeroexplicit}
\end{equation}
The momenta $p, q$ must be on mass shell since we are considering a classical theory, so we have
\begin{equation}
Cosh(p_0/\kappa)-e^{p_0/\kappa} \mathsmaller{\sum}p_i^2/ 2 \kappa^2 =Cosh(m/\kappa) \geq 1
\end{equation}
It is sufficient to show that $ f(p_0,p_i ,q_i)$ is always non-negative for timelike/null momenta on the mass shell. For  momenta $p, q$ in the region $p_0, q_0 \in[0,\kappa]$, the function $ f(p_0,p_i ,q_i)$ does not have a local minimum, and it can only reach its minimum when $p, q$ are null momenta and $q_0$ reaches the maximum value allowed, i.e. $\kappa$.  In this situation we have
\begin{equation}
\frac{\partial f(p_0,p_i ,q_i)}{\partial p_0} >0,\ \ \  \text{for}\  {p}_0>0
\end{equation}
Thus the function $f(p_0, p_i, q_i)$ only reaches its minimum when $p_0=0$, 
\begin{equation}
f(p_0, p_i, q_i) \geq f(0, 0, q_i)=0,\ \text{when}\ p_0, q_0 \in [0,\kappa]
\end{equation}

This means  $v^\mu  ( V^p_q )_\mu^0 >0$ when $v^\mu$ is timelike or null. Thus for right-oriented loops the left-hand side of Eq.(\ref{causal}) satisfies the following condition,
\begin{equation}
\sum_{i=1}^n \tau_{i,i+1} v_{i,i+1}^\mu  ( \prod_{i<j \leq n} H_{j} )_\mu^0 = \sum_{i=1}^n \tau_{i,i+1} v_{i,i+1}^\mu ( V^{p_{n,1}}_{p_{i,i+1}})_\mu^0 >0, \ \ if v_{i,i+1}^0>0\ \forall i
\end{equation}
This tells us that the chain product of right transport operators cannot reverse a future-pointing four-velocity to past-pointing.  Hence right-oriented loops will not form a causal loop.

We now analyze the reason why the set of nonorientable loops includes a lot of causal loops. For the nonorientable loops, which have effective curvature $H_{tot}- \identity \neq 0$, the right-hand side of (\ref{causal}) does not vanish. Regardless of whether or not $ ( \prod_{i<j \leq n} H_{j} )$ can reverse the future-pointing $v_{i,i+1}^\mu$ to past-pointing,  it is easy to find a specific point $x$ on the cotangent space that satisfies   
\begin{equation}
\sum_{i=1}^n \tau_{i,i+1} v_{i,i+1}^\mu  ( \prod_{i<j \leq n} H_{j} )_\mu^0 = x_{1_{[ p_{n,1} ] }}^\nu \left(\identity-H_{tot}\right)_\nu^0>0, \ \ \  v_{i,i+1}^0>0\ \forall i
\end{equation}
Thus the nonorientable loops which are x dependent and have effective curvature can violate causality. The type of loops which was studied in \cite{Chen:2012fu} is causality violating exactly because of the above analysis.

\subsection{Orientable loops preserve global momenta conservation}
In Ref.\cite{Banburski:2013wxa}, it was shown that in relative locality some classical loops can break global momenta conservation with all the vertices safisfying local momenta conservation, due to the noncommutativity or nonassociativity of addition rule. Similar phenomena also occur in $\lambda \phi^4$ theory on $\kappa$-Minkowski spacetime, in which violations of energy-momentum conservation are to be expected in any given particle-propagation process because of the nonplanar diagrams in the loop correction\cite{AmelinoCamelia:2001fd}.

As we know, in curved spacetime there is no total momentum conservation in general. Relative locality was proposed as a dual picture of general relativity, so we may expect that global momenta conservation may be broken if the loop process is not flat. In this section, we will show that this is indeed the case. In the previous sections, we have established that the flatness of loops is equivalent to their orientability.  Here we will show that, for any noncommutative momentum space (not only for $\kappa$-Poincar\'e), orientability of loops leads to global momenta conservation, while for nonorientable loops the globle conservation will be broken.
\\ \\
For the right-oriented loop, the conservation of momenta on the vertices are given by
\begin{equation}
\mathcal{\tilde{K}}_i=l_i \ominus p_{i,i+1} \oplus p_{i-1,i} \equiv 0
\end{equation}
where $l_i$ are all the external momenta that enter the loop through vertex i. Solving the above equation for $l_i$ (we can solve it out because the right inverse property (\ref{ip}) is satisfied), one can immediately get that the sum of all the external momenta around the whole loop is
\begin{equation}
\sum^{n}_{i=1}{\oplus l_i} = \sum^{n}_{i=1}{(\ominus p_{i-1,i} \oplus p_{i,i+1})}\equiv 0
\label{conserveright}
\end{equation}
Thus the right-oriented loop has global momenta conservation. Similarly for the left-oriented loops, the conservation law on the vertices are
\begin{equation}
\mathcal{K}_i=l_i\oplus p_{i-1,i}\ominus p_{i,i+1} \equiv 0
\label{conservationlaw}
\end{equation}
Solving for the external momenta $l_i$ in the above equation using  the left  inverse property (\ref{ip}), the sum of the external momenta is
\begin{equation}
\sum^{n}_{i=1}{\oplus l_{n-i+1}} = \sum^{n}_{i=1}{(\oplus p_{n-i+1,n-i+2} \ominus p_{n-i,n-i+1})}\equiv 0
\label{conserveleft}
\end{equation}
Thus we get that orientable loops, i.e. flat loops have global momentum conservation. Note here that the order of summation of the external momenta is unique (up to cyclic permutations). For the right-oriented loops, the total global momenta Eq.(\ref{conserveright}) have to be summed in an order that has a right orientation. Similarly for left-oriented loops, the total summation Eq.(\ref{conserveleft}) to be summed in an order that has a left orientation. An arbitrary order of summation of external momenta will not lead to any conservation. This is a new notion of conservation and a more subtle condition that is different from usual linear momentum space.
\\ \\
Now let us consider nonorientable loops. We will show that the above statements do not hold in this case. To construct a nonorientable loop, we can take the above right-oriented loop and switch the m'th vertex to be left oriented. Now the loop has one vertex whose orientation is opposite that of the other vertices. The global momenta of the whole loop will be
\begin{equation}
\sum^{n}_{i=1}{\oplus l_i} =\ominus p_{n,1}\oplus (p_{m-1,m} \oplus p_{m,m+1}) \ominus (p_{m,m+1} \oplus p_{m-1,m})\oplus p_{n,1} \neq 0
\label{break}
\end{equation}
Only commutativity or a special fine-tuning of momenta values can make the above equation to zero. It is straightforward to generalize this simple argument to a loop that has more than one vertex whose orientation is opposite that of the other vertices. Such loops do not have global momenta conservation.

What about the other type of nonorientable loops? We know that the only other nonorientable loops must have nonorientable vertices. We can construct a simpliest example by changing the m'th vertex to be nonorientable in the above right-oriented loop i.e. if the conservation law is in the form of Eq.(\ref{nonorientable}):
\begin{equation}
\mathcal{K}_m=l_m\ominus p_{m,m+1}\oplus l'_m \oplus p_{m-1,m} \equiv 0
\end{equation}
we will get the following equation,
\begin{equation}
\sum^{n}_{i=1}{(\oplus l_i)}  \ominus p_{n,1} \oplus l'_m \oplus p_{n,1} =0
\end{equation}
Thus there is no order of summation that leads to global momenta conservation for this kind of loops. It is also straightforward to generalize this analysis to a loop that has more than one nonorientable vertex. 

In the above situations, the momenta external to the loop can only be conserved without fine-tuning, when the momentum space does not have torsion. For nonorientable loops, which have effective curvature due to the chain of nonlinear interactions, global momenta conservation will not hold.  This proof also supports the previous observation in the study of the one-loop correction of the proporgator in $\lambda \phi^4$ theory on $\kappa$-Minkowski spacetime \cite{AmelinoCamelia:2001fd}.

Note that the discussion above is restricted to associative addition. In the nonassociative case in general there is no global momentum conservation. We can easily see this because even Eq.(\ref{conserveright}) and (\ref{conserveleft}) will not be zero if the addition law is not associative.

\subsection{Loop compositions}
\begin{figure}[h]
  \centering
    \includegraphics[width=0.4 \textwidth]{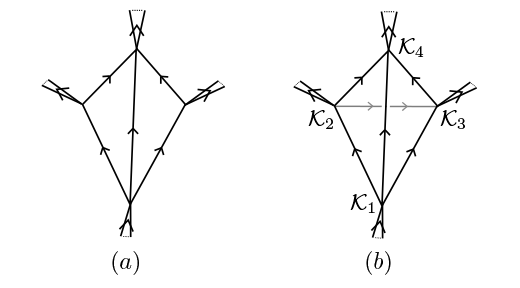}
	\caption{ Planar event-net (a) and nonplanar event-net (b) given by composition of single loops}
\label{planar}
\vspace{0 mm}
\end{figure}

The proofs in the last section are for single-loop processes. Here we briefly discuss what will happen for the processes that  are composed of  a few loops.  Similarly to the notion in  quantum field theory \cite{'tHooft:1973jz},  we say that a piece of event-net is {\textit{planar}} if it can be embedded in a two-dimensional plane in such a way that no two edges meet each other except at a vertex to which they are incident, e.g.[\ref{planar}] (a). If it cannot be embedded in a two-dimensional plane without edges crossing, we say the piece of event-net is  {\textit{nonplanar}}, e.g. Fig.[\ref{planar}] (b).

If a few single loops are composed into a planar event-net, all of the properties that were proved in the previous sections can be easily generalized.  If there are orientable loops that share edges with each other [an example is shown in fig.[\ref{planar}](a)], all single loops connected by the edge would have to have the same orientation, and we will have flatness, x independence, causality perservation, and global momentum conservation of the whole event-net when the momentum space can be considered as $\kappa$-Poincar\'e.  If  there are orientable loops in the planar event-net which just share vertices with each other but not edges, they can have the opposite orientations to enjoy all the properties that an orientable single loop has. We can understand this by the fact that the orientation of a vertex is defined relative to a loop -- external and internal momenta are relative when loops are composed.  However, in a piece of event-net, as long as there is one loop that is nonorientable, the whole event-net will be x dependent, and global momenta conservation will be broken as well.

If a few single loops are combined into a non-planar event-net, the situations are complex and we will not give a general proof here. A preliminary analysis implies that nonplanar loops are always x dependent and  they break global momentum conservation. We conjecture that in order for nonplanar loops to be x independent, we need a commutative and associative (i.e. flat) momentum space. We give a simple example of a nonplanar loop in [\ref{planar}](b). Though this kind of loops can be solutions of the theory with timelike/null on-shell momenta, there is no way to write the vertices to achieve x independence: for example consider vertex $\mathcal{K}_1$, it is shared by three single loops and $\mathcal{K}_1$ will be a nonorientable vertex in one of them.

\section{Conclusion and Discussion}
In this paper, we have classified single-loop processes by their orientation. We have shown that when momentum space is taken to be $\kappa$-Poincar\'e (or any associative momentum space),  nonorientable loops have effective curvature due to combinations of nonlinear interactions. These kinds of loops in general are x dependent; they can break causality and do not have global momentum conservation (without fine-tuning the momenta value). Orientable loops however are flat, x independent and have global momenta conservation. They preserve causality when the momentum space is taken to be $\kappa$-Poincar\'e. The relationships between the few properties are as follows:

\begin{table}[h]
\centering
\begin{tabular}{ l l c r r } 
 Causal\ \  \ \ & $\Leftarrow$ &The loop is orientable& $\Leftrightarrow$ &\ \ x independent \ \  \\
   &  &$\Updownarrow$ & &\ \ \  (the loop is flat) \\
   &  & Global momenta conserved  & &  \\
\end{tabular}
\end{table}

Does this mean that we should rule out nonorientable loops since they have weird properties? Actually we have not found fundamental reasons for choosing some kinds of vertices rather than others. In the effort of constructing a quantum field theory of relative locality, we need to consider all kinds of vertices in principle.  All the analysis (apart from the causality) in this paper is also true for off-shell loops. We can expect that nonorientable loops will lead to challenges in constructing a quantum field theory, for example, the lesson from quantum field theory in curved spacetime shows that unitarity is lost if there exist closed timelike curves\cite{Friedman:1992jc, Boulware:1992pm}.  However, as we have already discussed, nonorientable loops have effective curvature. In general relativity, the holonomy around a loop of geodesics depends on the momenta, since geodesics are generated by four-velocity. Hence it is not surprising that in a nonlinear momentum space, a loop is dependent on variables conjugate to the momenta. Moreover, in curved spacetime, there is no global momenta conservation in general and closed timelike curves can be solutions of the theory. The nonorientable loops have similar properties. 

It is important to point out that all of these strange effects are only apparent when the energy scale is close to $\kappa$   \cite{Chen:2012fu},\cite{Banburski:2013wxa}, which is assumed to be the Planck energy. Future research should look for testable predictions of the weird nonorientable loops. One way might be looking at whether these small $1/\kappa$ effects can be amplified in statistical properties in astrophysical situations. Another avenue might be exploring the x dependence. In a big event-net, as long as one of the loops is x dependent, the whole event-net becomes x dependent. In this case, the $x$ on the cotangent spaces behave like ``hidden variables" that are attached to the event-net. The prediction of the theory is then that even in the classical case, such nonorientable loops are not freely repeatable in experiments (with exactly the same momenta, vertices, and  proper time of propagation).  However, with the requirement of precision at the $1/\kappa$ level, quantum uncertainty becomes dominant. Thus we need a quantum theory of relative locality to draw solid conclusions. It remains to be seen whether the weird nonorientable loops are to be ruled out, or whether they are new features in phenomenology that are an embodiment of the Born reciprocity principle.

\vspace{5mm}
{ \bf Acknowledgements:}
I am very grateful to A. Banburski, L. Freidel  and L. Smolin for helpful discussions and suggestions. Special thanks to G. Amelino-Camelia for encouraging this research,  and sharing the results of his research group. This research was supported in part by Perimeter Institute for Theoretical Physics. Research at Perimeter Institute is supported by the Government of Canada through Industry Canada and by the Province of Ontario through the Ministry of Research and Innovation. The work was supported by NSERC and FQXi.

\end{document}